*Bioinformatics Advances (to appear)*

# Phil Bourne (1953–2026): From Small Molecules to Big Data — *The Journey of a Multifaceted Visionary*

Andreas Prlić[1], Cameron Mura[2*]

[1] Clinical Genomics Engineering; Natera; USA; https://orcid.org/0000-0001-6346-6391

[2] Department of Biomedical Engineering and School of Data Science, University of Virginia; Charlottesville, VA; USA; https://orcid.org/0000-0001-7985-2561

* Corresponding author. Email: cmura@virginia.edu

## Abstract

Born in London in 1953 and raised in Australia, Phil Bourne spent over four decades in science moving across scales: from crystal structures to the world's premier structural biology database; from scientific journals to national data policy; from molecules to institutions. The International Society for Computational Biology (ISCB) chose Phil to receive its 2026 Outstanding Service Award (Wiper, 2026). He passed away on 8 March 2026, in Charlottesville, Virginia, before he could receive it. This piece is in memory of Phil, who most recently served as the Founding Stephenson Dean of the University of Virginia's School of Data Science and as a Professor of Biomedical Engineering—and who was, over the course of his career, a steadfast builder of Computational Biology and Bioinformatics, of Open Science and Data Science. A tribute site, https://philbourne.muchloved.com, invites those who knew Phil to share memories, photos, and reflections.

## Early Life & Education

"*Phil loved and identified strongly with all three countries that formed him: England, his birth; Australia, his education; and the USA, his family and his career.*"

— Tony Fletcher, Phil's oldest friend, speaking in a March 2026 memorial service (UVA-DataSci-Memorial, 2026)

Philip Eric Bourne was born in London on 22 March 1953 and raised in Adelaide, Australia from his teen years onwards. Despite being discouraged from the subject by his high school science teacher, Phil chose to follow his passion and elected to study chemistry at Flinders University, going on to complete a PhD in physical chemistry there in 1979. His dissertation concerned X-ray crystal structures of metal complexes (Figure 1; Bourne & Taylor, 1980), and in the process of conducting that research he began writing computer programs to make sense of the crystallographic data. After Flinders, Phil moved from small molecules to the young field of protein crystallography, and that era's structural biology world was smaller than one might think. After a postdoc at Sheffield with Pauline Harrison (Andrews et al., 2024), working on ferritin (Bourne, 2017), Phil arrived in the U.S. on Thanksgiving 1981 to join Barbara Low's laboratory at Columbia; Low and Harrison had both trained under Dorothy Hodgkin at Oxford. Low had also been Helen Berman's mentor, which was a connection that mattered when Phil and Helen's paths converged, a decade later, over the future of the Protein Data Bank (PDB).

## Structural Biology: The PDB Years

"*Writing and running computer programs that defined the positions of atoms in matter was so cool, I was hooked… The thrill of mapping out atomic structures that no one had seen before remains with me to this day.*"

— Phil Bourne, writing in *PLOS Biology* in 2017 (Bourne, 2017)

Through the late 1980s and into the 1990s, as personal computers became increasingly important scientific instruments, Phil's focus shifted from crystallography to computation. He joined the faculty at the University of California San Diego in 1995 and, almost immediately, his attention turned toward a question that would define the next chapter of his career: what should the infrastructure of structural biology look like in the coming age of the internet and large-scale data?

The answer lay in the Protein Data Bank (PDB), the comprehensive global archive where structural biologists—crystallographers, NMR spectroscopists, and eventually cryo-electron microscopists—deposit their work. Established at Brookhaven National Laboratory in 1971 (Berman, 2008) with just seven protein crystal structures, by the mid-1990s the PDB had grown to thousands of entries yet faced an uncertain institutional future. In 1998, Phil joined forces with Helen Berman at Rutgers and Gary Gilliland at NIST to form the Research Collaboratory for Structural Bioinformatics (RCSB), which took over management of the PDB in 1999. Phil led the UCSD site, responsible for the web interface, query systems, and structure visualization tools. Peter Arzberger served alongside Phil at UCSD.

Under RCSB stewardship and as part of the Worldwide PDB (wwPDB), the PDB grew from roughly 10,000 structures in 1999 to well over 250,000 by the time of Phil's death. When DeepMind's AlphaFold 2 demonstrated in 2020 that AI could predict protein structures with experimental accuracy (Jumper et al., 2021), the training data was the PDB. Phil had spent two decades helping build that foundation.

With over 110,000 citations across his career (Bourne-Scholar-Profile, 2026), Phil's scholarly impact was unequivocally substantial. Nevertheless, Phil himself questioned what such citation metrics truly signify. In a 2008 perspective piece entitled "I Am Not a Scientist, I Am a Number", Phil noted that his most-referenced paper had been cited thousands of times, "*but he suspects hardly anyone has ever read it; it is a reference to a commonly used database he helped develop*" (Bourne & Fink, 2008). This is but one example of Phil thinking carefully about how science should be done (as a practice and as a profession), and sharing those thoughts clearly—by his own example, by direct mentorship, and in print via his lucid writings.

Phil understood early that a database nobody can use is just storage. Under his and Helen Berman's co-directorship, the RCSB PDB developed data standards, most notably the mmCIF format that Phil spearheaded with John Westbrook and others (Westbrook & Bourne, 2000). Those standards later fed directly into the FAIR principles (Findable, Accessible, Interoperable, Reusable), which Phil co-authored in 2016 (Wilkinson et al., 2016) and which are now the standard framework for scientific data infrastructure.

The UCSD era also saw a major effort in immunoinformatics. Phil co-founded the Immune Epitope Database (IEDB), established in 2003 with NIH funding and described in 2005 (Peters et al., 2005). With Julia Ponomarenko leading the UCSD effort at SDSC, Phil helped build the database's structural arm (Ponomarenko et al., 2011), linking curated epitope data to PDB coordinates, and he co-authored ElliPro, a structure-based tool for predicting antibody epitopes (Ponomarenko et al., 2008). The IEDB now catalogs more than 1.6 million immune epitopes and remains one of the field's essential open resources.

Beyond the PDB and immunoinformatics, Phil's lab at UCSD pursued a wide range of research over nearly two decades. His interests and directions included structural bioinformatics, systems biology, *in silico* pharmacology and drug discovery, cell signaling, and a sustained thread on molecular evolution, e.g., as regards fold space (Shindyalov & Bourne, 2000; Xie & Bourne, 2008), the natural history of kinases (Scheeff & Bourne, 2005), and spliceosome evolution (Veretnik et al., 2009). Phil's influential *Structural Bioinformatics* text (Bourne & Gu, 2009) went through two editions, co-edited with Helge Weissig and then Jenny Gu. The Combinatorial Extension (CE) algorithm, developed with Ilya Shindyalov in 1998, aligned protein three-dimensional structures by building from fragment pairs based on local geometry (Shindyalov & Bourne, 1998). It became a widely popular and standard structural biology tool.

Phil was a deeply collaborative scientist who worked with dozens of researchers across these many areas, at UCSD and beyond (Figure 2). The work mentioned here is only a fraction of his group's output, and numerous colleagues, students, and other contributors are not individually named here simply because of space constraints (Bourne-Lab-Alumni, 2022 supplies a curated library of previous members).

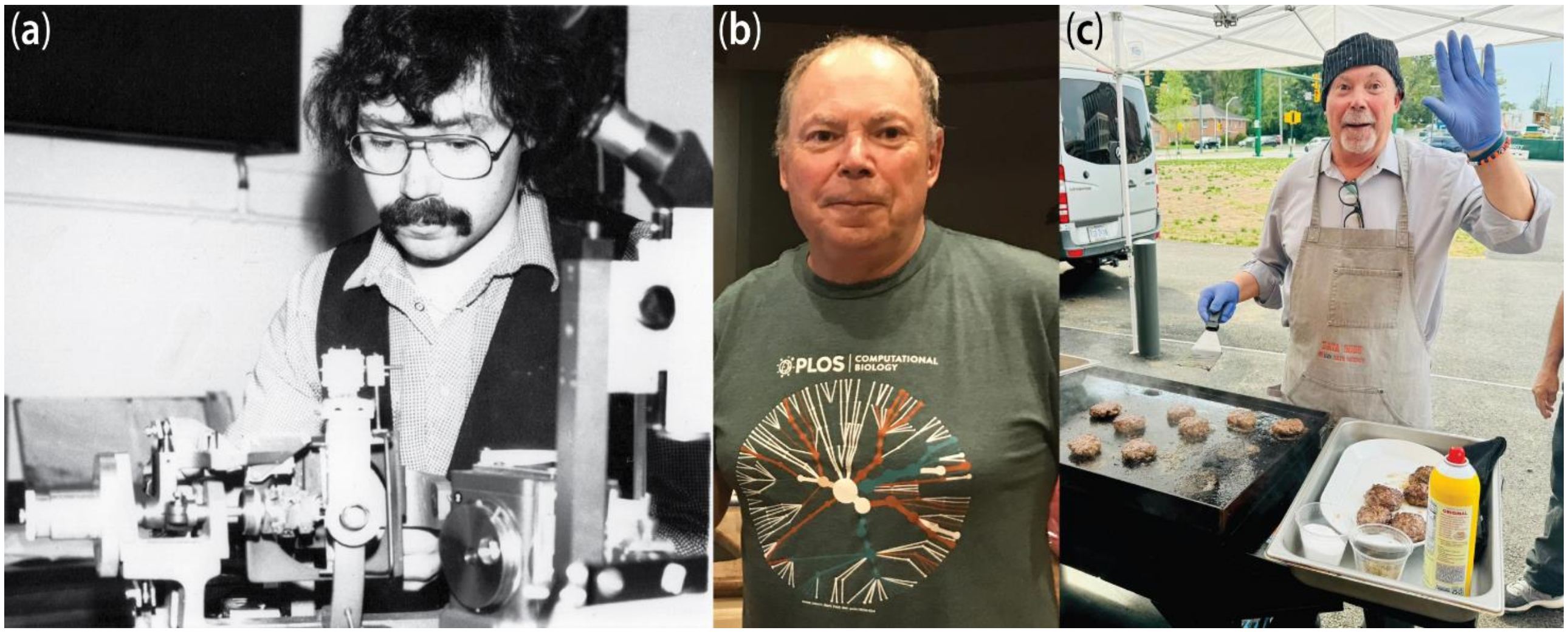

Figure 1. Phil in action: (a) performing X-ray diffraction studies with a precession camera in the late 1970s, (b) at home in Charlottesville in Fall 2018, and (c) in Fall 2024, at an annual 'Phil-on-the-Grill' kickoff event at UVA's School of Data Science. Don Brown remarked that "*I think all of us remember Phil-on-the-Grill. I mean, what a guy. He would flip hundreds of hamburgers in order to just be part of the community.*"

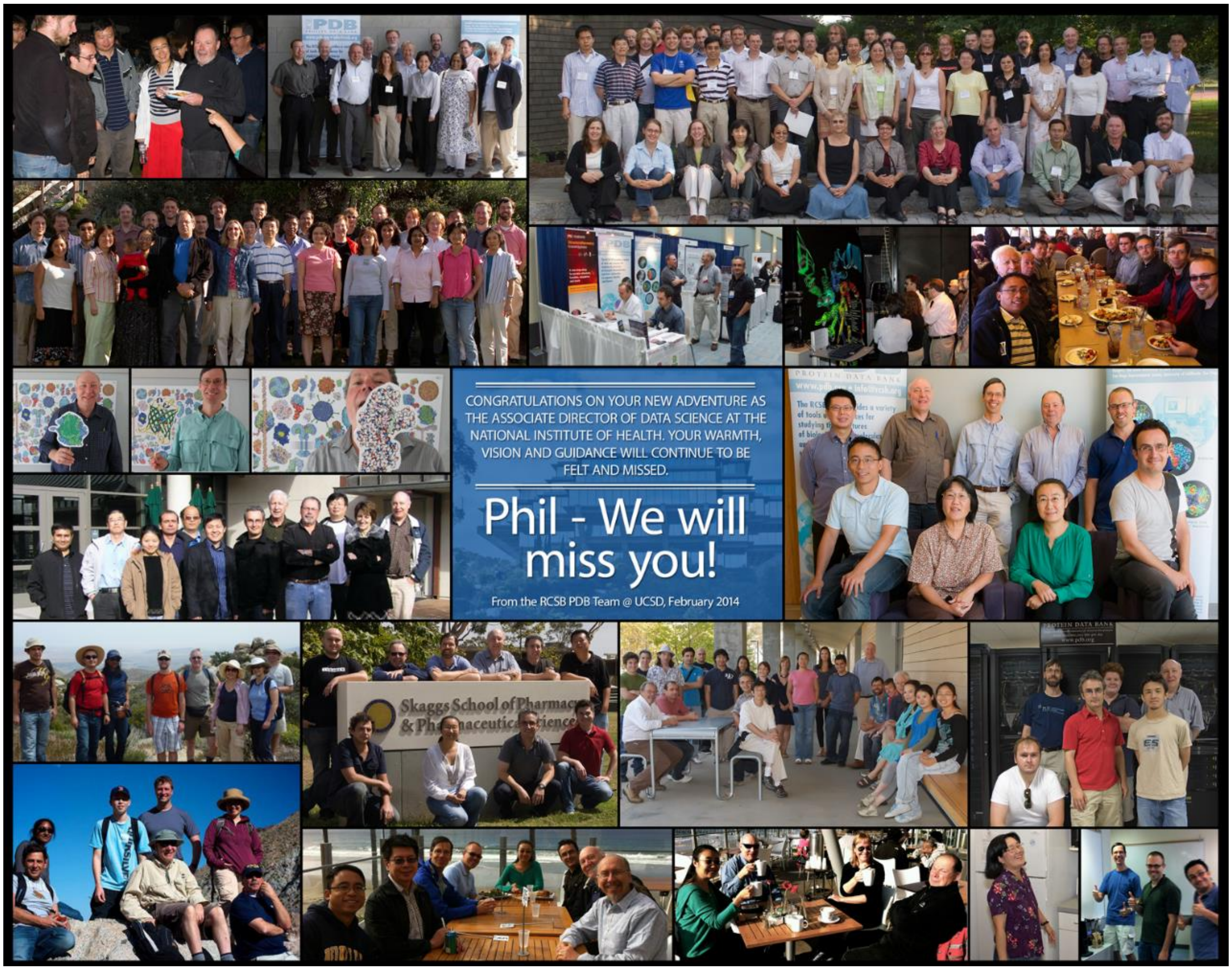

Figure 2. A goodbye card for Phil, upon his departure from UCSD and move to the NIH, depicting many former lab members and collaborative partners from the early 2000-2010s.

## From Computational Biology to Open Access: A Broader Vision

"*[Phil] was an inspiration; a radical; a fierce advocate of open science and data; and a fun colleague.*"

— Janet Thornton, writing in ISCB's Newsletter (ISCB-News-Mar2026)

Protein structures were not an end in themselves for Phil. Rather, they supply a foundation for understanding biology at every scale, from molecular interactions to population-wide dynamics: "Life is three-dimensional, and it begins with molecules," as he wrote in *PLOS Biology* in 2017 (Bourne, 2017). To Phil, the centrality and key significance of molecular/structural information was all that much more reason it should be freely accessible. As cogently articulated by Iratxe Puebla of FORCE11 (Puebla, 2026), Phil lived by the mantra "*Complete access to all knowledge by everyone on the planet, and the ability to use that knowledge in new ways that only those people can imagine.*" To Phil, making invaluable structural data truly open was a prerequisite to future progress in the life sciences, both basic and applied.

These convictions made Phil an early advocate for open-access publishing. In 2005 he co-founded *PLOS Computational Biology* and served as its first Editor-in-Chief for eight years. Conceived of as "a new community journal" (Bourne et al., 2005)—accountable to the field and its scientists, rather than a commercial publisher—the journal set a standard for open, community-oriented publishing that the discipline still measures itself against.

Throughout his career, Phil kept finding new ways to lower barriers in scientific communication. In 2007 he co-founded SciVee, a "YouTube for science" that paired peer-reviewed papers with short, researcher-narrated videos. The premise was simple: knowledge kept behind paywalls, or buried in dense prose, was knowledge lost.

As recounted in a 2023 tribute (Mura et al., 2023), Phil's advocacy extended well beyond journals, and was recognized in 2009 by the *Benjamin Franklin Award* for promoting free and open access to the materials of science (SDSC-News-Apr2009) and, in 2010, by Microsoft's *Jim Gray Award for eScience* (Microsoft-News-2010).

In a continuation of the idea that knowledge relegated to PDFs is knowledge obscured, Phil hosted a "Beyond the PDF" workshop in 2011, with the goal of "moving digital research communications beyond putting digital paper online". This workshop would lead Phil to co-found FORCE11, the Future of Research Communications and e-Scholarship, an international community aimed at modernizing how scholarship is created, shared, and credited (Puebla, 2026). As both an idealist and a realist, Phil took great pains to show that openness paid: a 2016 essay in *eLife* that he co-authored marshaled evidence that sharing data, code, and preprints tends to advance the individual scientist's own career, not only the common good (McKiernan et al., 2016). Openness, Phil argued, is not a charity that science affords in good times, rather it is the condition under which science works best.

Fundamentally, these endeavors—the PDB, *PLOS Computational Biology*, SciVee, FORCE11—reflected the same basic principle in different guises: Science advances optimally when it is open, connected, and accessible. Phil had been making that argument since the mid-1990s, well before it became commonplace or consensus. And underneath all the infrastructure, the core principle was always about people: Every barrier Phil could lower would let someone else in.

## Science Communication & Mentoring: Ten Simple Rules

"*When you are long gone, your scientific legacy is, in large part, the literature you left behind and the impact it represents. I hope these ten simple rules can help you leave behind something future generations of scientists will admire.*"

— Phil Bourne, writing in the inaugural *Ten Simple Rules* (Bourne, 2005)

Phil's computational biology and open-access contributions alone would have comprised a distinguished career. And yet another chapter began almost as a side project.

In October 2005, alongside the launch of *PLOS Computational Biology*, Phil published a two-page editorial entitled "Ten Simple Rules for Getting Published" (Bourne, 2005). The response was enthusiastic enough that he wrote a second piece, and then a third. The series grew to cover collaboration, grant writing, poster presentations, graduate school, postdoctoral selection, reputation building, starting a company, academic promotion, and even how to win a Nobel Prize. Phil's collaborators contributed their own Rules, and eventually the series opened to the wider community. By December 2018, when Phil co-authored "One Thousand Simple Rules" (Bourne et al., 2018)—the hundredth article in the series, 13 years after the first—the collection had been translated into at least six languages and drawn many millions of views. Why has this series resonated so strongly and widely, and for so long? Perhaps because it became a place where Phil (and then others) could plainly describe the unwritten rules of scientific and professional life, and offer practical guidance—a forum to "...empower scientists to more effectively navigate the world of very-human scientific activities (papers, talks, careers) that begin where the data-collection and number-crunching end" (Mura et al., 2023).

Read in sequence, the Rules let you track Phil's career: Each piece was written from where he stood at the time. For example, Phil's 2019 piece, "Ten Simple Rules to Aid in Achieving a Vision" (Bourne, 2019), reads more like a memoir than a manual. Rule 10, "Take time to enjoy the experience", describes celebrating milestones with collaborators over dinners, "*sharing the creative experience through social activities with families.*" The series eventually tracked Phil to UVA, with 2022's "Ten Simple Rules for Good Leadership" (Bourne, 2022). Phil's final TSR article, 2023's "Ten Simple Rules for Humane Data Science" (Masum & Bourne, 2023), includes rules like "*Help people live better lives*" and leverage algorithms to "*raise the floor*". Whether dismantling barriers via open access or building connections between people, Phil had a gift for raising the floor. For, as Berman & Zardecki (2026) have pointed out, Phil "*followed his own rules.*"

## Building FAIR Data Science at the National Institutes of Health

"*Phil saw biological data not as separate collections belonging to separate groups or fiefdoms, but a commons—a shared resource that belonged to everyone through science that everyone could use. That... seeded a lot of the conversations about... biomedical data programs at NIH.*"

— Vivien Bonazzi (NIH colleague and friend), speaking in a March 2026 memorial service (UVA-DataSci-Memorial, 2026)

In 2014, Phil accepted a role that had not previously existed: Associate Director for Data Science at the NIH, where he also served as a Senior Investigator at the National Center for Biotechnology Information. His mandate was to lead the $110M Big Data to Knowledge (BD2K) initiative, building the training, tools, and data infrastructure that biomedical research would need as data generation scaled beyond any single institution's capacity to manage it (Bourne et al., 2015).

Phil brought the FAIR principles he had helped formulate as a guiding framework, pushing NIH to treat data as a primary research product rather than a byproduct. He championed data sharing through a shared commons, advocated for persistent identifiers for research objects (data, software, publications) and piloted researcher access to commercial cloud computing before most federal agencies had a policy for it. Using the challenge grant mechanism, Phil also helped launch the Open Science Prize (Kittrie et al., 2017), an international public-private effort to encourage the creative reuse of open data across disciplines.

While breaking down silos in an institution of NIH's scale proved relatively slow, an experience Phil later refracted through Rule 9 of "Ten Simple Rules to Aid in Achieving a Vision": *Know when to stop* (Bourne, 2019), the agenda he set has come to fruition—in the Common Fund Data Ecosystem, the NLM Data Set Catalogue, the STRIDES cloud-access initiative, and NIH's DataCite partnership for persistent identifiers across biomedical data resources. He mentored NIH staff and helped grow leaders who have carried that vision forward.

Phil left the NIH in 2017 for the University of Virginia, where there arose a new opportunity to build data science from the ground up.

## UVA, A School Without Walls, and Biomolecular Data Science

"*What an incredible scientist, mentor, and leader. I feel blessed to have had the chance to work with him and learn from him.*"
— Shayn Peirce-Cottler, UVA Biomedical Engineering colleague

"*He always seemed to spark some new way to think about a problem and to do it in a way that often brought a smile to my face.*"
— Don Brown, UVA Data Science colleague

Phil came to UVA in May 2017, joining its Biomedical Engineering department and taking over as director of a still-young Data Science Institute (DSI), which had been led by Don Brown of UVA's Systems Engineering (UVA-DataSci-Story, 2026). By September 2019, Phil had shepherded the DSI into a new School of Data Science (SDS)—the twelfth school in UVA's 200-year history, and the first autonomous school of data science in the United States (Moody, 2019). As its founding Stephenson Dean, Phil called this new unit a "*School Without Walls*": Deliberately without departments, its faculty seeded across the sciences, engineering, medicine, social sciences, humanities, and public life. The School grew from a handful of personnel to over 100 faculty and staff by 2025, weathered being founded six months before a historic pandemic, and gained a building of its own in 2024, funded by the largest private gift in the University's history (Hester, 2019). Expanding upon a vibrant MSDS (MS in Data Science) program that began in 2014's DSI, Phil led the School in its launch of new educational programs in the 2022-2025 timeframe, including the first doctoral degrees conferred in Spring 2024 and the first undergraduate majors enrolled that Fall. UVA President Emeritus Jim Ryan commented on Phil's deeply prescient leadership in a March 2026 memorial service (UVA-DataSci-Memorial, 2026), stating "*Phil saw things that others didn't. He knew what was coming in data science... and how it was already changing the world within and outside of academia and would continue to do so.*"

Phil partnered with one of us (CM) in 2018 to begin a *Biomolecular Data Science* lab, and the science there followed the same tenets as the School: Collaboration, curiosity, and no walls. Phil's UVA research continued some threads from UCSD while opening new ones. For example, his group extended the lab's long-running kinase investigations by pursuing structural profiling of the human kinome, with implications for drug selectivity and resistance (Zhao & Bourne, 2023, 2026). Another line of work devised an 'Urfold' model for protein structural relationships (Mura et al., 2019) and used it to explore fold space via deep generative modeling (Draizen et al., 2024), asking whether protein families form communities across a continuous landscape rather than fall into the discrete bins of traditional structural classification schemes (Bourne et al., 2022). A third direction, as part of a longstanding collaboration with Robert Preissner's team at Charité, turned to population-scale data science: mining electronic health records for real-world evidence, Phil and colleagues found that estradiol is associated with milder COVID-19 symptoms in women (Seeland et al., 2020)—a result that drew international coverage and helped inform a clinical trial. As at UCSD, these projects were pursued with many collaborators, and again many of them go unnamed here because of space constraints.

Throughout the UVA years, Phil continued pressing the open-science argument at the level of policy. The FAIR principles that he co-authored in 2016 anchored a run of position pieces on data governance, research software, and what biological databases must become in the age of generative AI—including how to sustain them. Indeed, another conspicuous aspect of Phil's approach to open-data and scientific governance was the issue of sustainability (of databases, knowledgebases, and other infrastructural resources). For example, in the last couple of years Phil worked with UVA economist Terry Johnson to formulate and analyze a "data sustainability paradox" (Johnson & Bourne, 2023), the idea being that the biological data resources that the field now depends upon are precisely the ones hardest to keep funded. (Incidentally, Phil also liked to point to that work as an example of the sort of productive interactions that may have never occurred in more conventional academic settings, e.g. a school *with* walls.) Phil's latest contribution in the data-governance realm appeared in *Science* on 5 March 2026 (Haendel et al., 2026), three days before he died. To the end, Phil was making the case that data is too invaluable to not belong to everyone.

Phil Bourne had many sides, as a scientist and as a human. He began with small-molecule crystal structures and ended with a School of Data Science. In between, one finds 45 years of scientific contributions and articles, research mentoring, the PDB, *PLOS Computational Biology*, the FAIR principles, industrial alliances and business ventures, and hundreds of 'simple rules'—and, somehow, annual multi-week motorcycle trips. Phil was fond of 'multiscale science', which is a phrase that could also describe his own mind: as comfortable with atoms and protein dynamics as with institutions and governance, as ready to think in nanoseconds as in decades or even geological timescales (Dupont et al., 2010). In a sense, the *School Without Walls* was the institutional expression of Phil's range. It made room for collaborations that a conventional structure could not. Emblematic of his leadership and vision, in recent years Phil conceived of UVA's 'Futures Initiative' to ready the University for the AI era (Bourne et al., 2023), and he set in motion a second data science building, oriented toward entrepreneurship (it was approved this June (Mather, 2026)). Phil achieved all of this as a dean who insisted that his office be no larger than any other faculty member's.

## Phil-the-Human: Humility, Openness, Generosity

"*Phil was a friend to all in a sincere and understated way... He took his work seriously and he knew it was important but he had no trace*

*of self-importance... Phil Bourne was as beloved a man as I have ever met and deservedly so.*"

— Jim Ryan, UVA President Emeritus, speaking in a March 2026 memorial service (UVA-DataSci-Memorial, 2026)

Phil is remembered as a brilliantly insightful scientist, selfless colleague, generous mentor, creative visionary, and humble leader—one who placed service to others and the common good above all, and whose influence will continue to shape bioinformatics, computational biology and data science for years to come. But who was he as a human? One may be tempted to take Phil's citation counts or professional accolades as a proxy for who he was. That would be a mistake, and an ironic one: Phil did not view one's worth in terms of awards, citation metrics, or other such measures of 'success'. We believe the key is that the qualities that made Phil a good scientist and a good human were one and the same (Mura et al., 2023): The openness he built into databases and journals was the same openness at the dinner table, and the generosity that gave structural bioinformatics and other fields their infrastructure sprang forth from the same generosity that would lead Phil to give a struggling colleague his time.

Phil was a major presence in every sense—physically present and focused in meetings, curious about everything, and happiest in motion or on-the-go. He flew planes and rode motorcycles across continents; he hiked, grilled burgers for students (Figure 1), traveled widely, and wrote about it along the way (Bourne-Bike-Blog, 2026). Michael Waterman, surveying all that Phil's research and service had given the field, remarked as many of Phil's colleagues did: "*But I will always think of him on his motorcycle!*" (ISCB-News-Mar2026). Commenting on Phil's natural curiosity and his passion for motorcycles, fellow biker and UVA colleague John Unsworth noted (UVA-DataSci-Memorial, 2026) that: "*why Phil loved motorcycling... had everything to do with curiosity... He just wanted to experience the world, the wind, and the wonder of the gyroscopic effect.*" Phil and Roma, his wife of 42 years, kept an open house wherever they lived, and to a great many people Phil was an encouraging mentor and a trusted confidant. Phil gave freely and openly to students and senior colleagues alike, with no apparent thought for what 'might be in it for him'. With Phil, human warmth wasn't secondary to the science, but rather its foundation. The same man who insisted that knowledge should belong to everyone couldn't help but extend that same openness to anyone with whom he came into contact. For these reasons, in considering what might be Phil's most lasting legacy, we would say '*people*'. Indeed, 'people', 'community' and related terms emerge as the most prominent themes across 14 separate tributes to Phil, spanning ≈ 90 minutes in a March 2026 memorial service at UVA's School of Data Science (Figure 3). Phil had a way of leaving people more capable, and more generous, than he found them—of getting, as Francis Ouellette put it, "*all of us to do a little bit more*" (ISCB-News-Mar2026).

Those lucky enough to have known Phil can attest that he wasn't just a fun and insightful scientific thinker, colleague and mentor, but also a compassionate individual. Indeed, some of Phil's most defining qualities—his authentic selflessness, graciousness and humility; his intellectual flexibility, curiosity and creativity; his expansive and forward-thinking perspective on subjects he got into; his unflinching positivity and optimism, tempered by a sense of pragmatism—have been beautifully recounted in several places since his passing. In one fitting tribute, mathematician Ken Ono, a UVA colleague who worked closely with Phil on the innovative *Futures Initiative*, wrote that "*To know Phil Bourne was to watch the future unfold... Beyond his brilliant mind was a boundless benevolence. A quiet, steadfast kindness. Phil possessed a rare and beautiful gravity. He had the vision to look over the horizon and... the heart to rally friends and colleagues.*" (Ono, 2026) How does one adequately capture such a human? — Perhaps via his oldest friend, Tony Fletcher, who knew Phil for 70 years and described him this way: "*He was a top bloke, a diamond geezer, like we say in London, and someone who's going to live on in everyone's hearts. It was really the privilege of my life to be his friend.*" (UVA-DataSci-Memorial, 2026) And, perhaps in Phil's own words: Asked near the end what he was most proud of (UVA-DataSci-Story, 2026), Phil did not name the school, the building, his science, or his citations. He answered simply: "*Everybody. Everybody at the School.*" These personal qualities are what "*made Phil Phil*" (to quote Vivien Bonazzi), shaping his legacy as a scientist, mentor, academic leader, visionary, and really—at the end of the day—just a good human being.

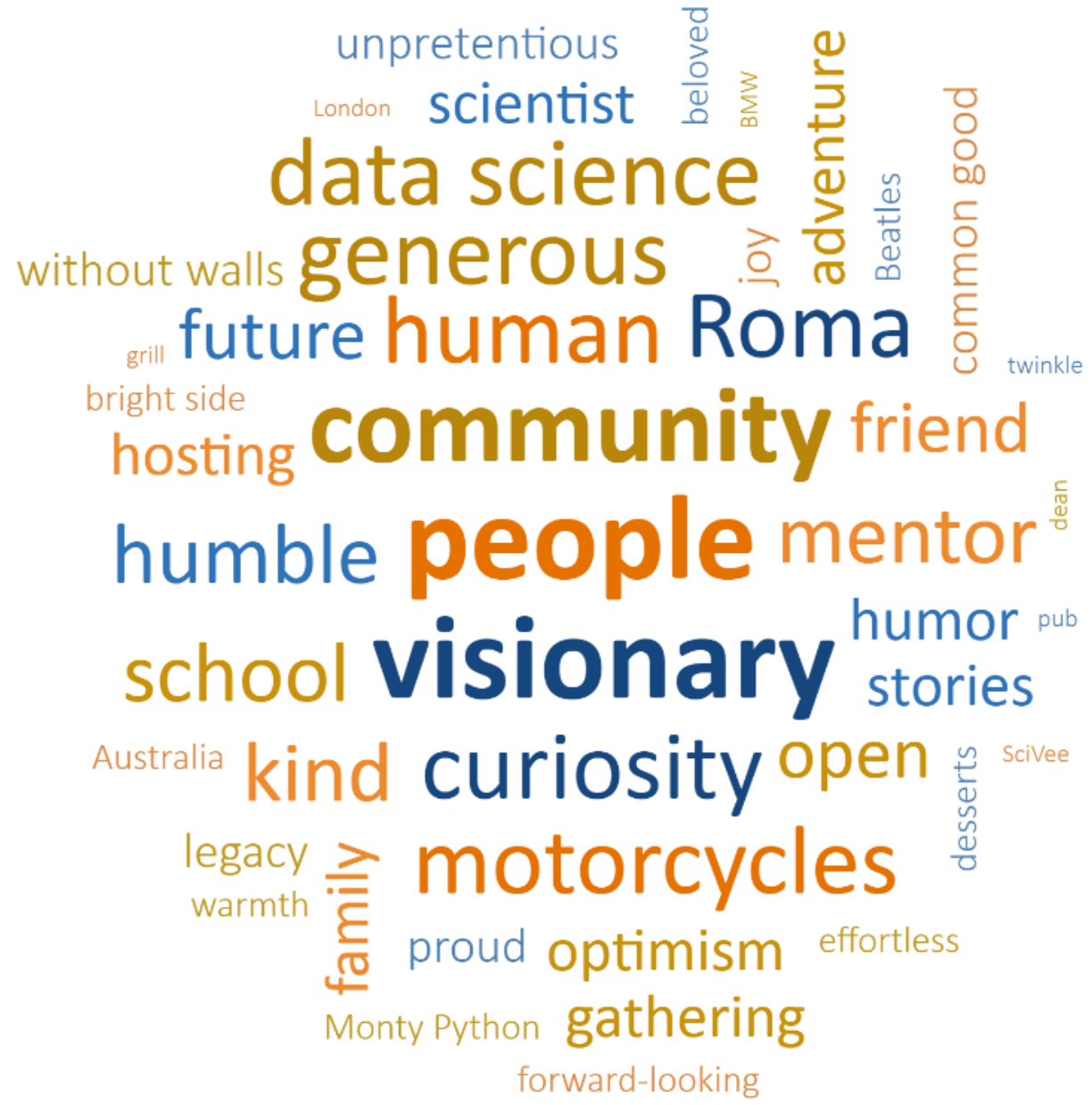

Figure 3. Recurring themes used to describe Phil, shown as a word cloud of the terms that statistically emerged in a memorial service held at UVA's School of Data Science on 20 March 2026. The cloud was generated in claude.ai from a transcript of the 14 speakers in the ≈90-minute service; terms are sized by how broadly a theme recurred across speakers and were semantically clustered, with synonyms and related word-forms collapsed into a single representative term. Word positions and orientations are arbitrary; colors are likewise arbitrary, simply reflecting two institutions where Phil spent much of his career (UCSD's blue and gold, and the blue and orange of UVA). Note the preponderance of terms relating to Phil's humanity, humility, and vision.

## Acknowledgements

This piece is dedicated to Phil's family: Roma Bourne (née Chalupa), his wife of 42 years; his daughter Melanie; and his son Scott & daughter-in-law Molly, along with their child Jessica, Phil's granddaughter. We thank Roma for feedback on the manuscript's biographical details, as well as colleagues at the NIH for contributing part of the text. We are grateful to Diane Kovats (ISCB) for making this project possible, and we thank Jenny Gu for reviewing an early version of the manuscript. Please note that this tribute provides only a glimpse of Phil's life in science; absence of citations reflects only space limitations and not opinions about Phil's many significant colleagues and coworkers over the years.

## References

Andrews SC, Theil EC, Harrison F *et al.* (2024) *In Memory of Pauline Harrison.* https://bioiron.org/news/in-memoriam/in-memory-of-pauline-harrison.aspx (11 August 2026, date last accessed).

Berman HM. The Protein Data Bank: A historical perspective. *Acta Crystallogr. A* 2008; **64**:88–95. https://doi.org/10.1107/S0108767307035623

Berman HM, Zardecki C. How Phil followed his rules. *Protein Sci.* 2026; **35**(5):e70577. https://doi.org/10.1002/pro.70577

Bourne-Bike-Blog. (2026) *Philip Bourne's TravelBlog.* https://www.travelblog.org/bloggers/ofl (11 August 2026, date last accessed).

Bourne-Lab-Alumni. (2022) *Previous [Bourne] Lab Members* https://docs.google.com/spreadsheets/d/1QZ48UaKcwDl_iFCvBmJsT03FK-bMchdfuIHe9Oxc-rw (13 August 2026, date last accessed).

Bourne-Scholar-Profile. (2026) *Philip E. Bourne on Google Scholar.* https://scholar.google.com/citations?user=Y9q2zZAAAAAJ&hl=en (11 August 2026, date last accessed).

Bourne PE. Ten simple rules for getting published. *PLoS Comput. Biol.* 2005; **1**:e57. https://doi.org/10.1371/journal.pcbi.0010057

Bourne PE. Life is three-dimensional, and it begins with molecules. *PLOS Biol.* 2017; **15**: e2002041. https://doi.org/10.1371/journal.pbio.2002041

Bourne PE. Ten simple rules to aid in achieving a vision. *PLOS Comput. Biol.* 2019; **15**: e1007395. https://doi.org/10.1371/journal.pcbi.1007395

Bourne PE. Ten simple rules for good leadership. *PLOS Comput. Biol.* 2022; **18**: e1010133. https://doi.org/10.1371/journal.pcbi.1010133

Bourne PE, Brenner SE, Eisen MB. PLoS Computational Biology: A new community journal. *PLoS Comput. Biol.* 2005; **1**:e4. https://doi.org/10.1371/journal.pcbi.0010004

Bourne PE, Draizen EJ, Mura C. The Curse of the Protein Ribbon Diagram. *PLOS Biology*. 2022; **20**(12): e3001901. https://doi.org/10.1371/journal.pbio.3001901

Bourne PE, Fink JL. I am not a scientist, I am a number. *PLoS Comput. Biol.* 2008; **4**(12): e1000247. https://doi.org/10.1371/journal.pcbi.1000247

Bourne PE, Gu J (eds). *Structural Bioinformatics (2nd ed).* Hoboken, NJ: Wiley-Blackwell, 2009.

Bourne PE, Lewitter F, Markel S *et al.* One thousand simple rules. *PLOS Comput. Biol.* 2018; **14**:e1006670. https://doi.org/10.1371/journal.pcbi.1006670

Bourne PE, Lorsch JR, Green ED. Perspective: Sustaining the big-data ecosystem. *Nature* 2015; **527**:S16–7. https://doi.org/10.1038/527S16a

Bourne PE, Ono K, Acampora C. (2023) *UVA Launches Futures Initiative to Chart Next Decade in Higher Ed.* https://datascience.virginia.edu/news/uva-launches-futures-initiative-chart-next-decade-higher-ed (11 August 2026, date last accessed).

Bourne PE, Taylor MR. The structure of aqua[3-ethoxy-2-oxobutyraldehyde bis(thiosemicarbazonato)]zinc(II). *Acta Crystallogr. B*. 1980; **36**(9): 2143–2145. https://doi.org/10.1107/s0567740880008151

Draizen EJ, Veretnik S, Mura C, *et al*. Deep generative models of protein structure uncover distant relationships across a continuous fold space. *Nat. Commun.* 2024; **15**: 8094. https://doi.org/10.1038/s41467-024-52020-2

Dupont CL, Butcher A, Valas RE *et al*. History of biological metal utilization inferred through phylogenomic analysis of protein structures. *Proc. Natl. Acad. Sci USA* 2010; **107**(23): 10567–10572. https://doi.org/10.1073/pnas.0912491107

Haendel MA, Ahern R, Bailey KB *et al*. Governing real-world health data as a public utility. *Science* 2026; **391**: 993–996. https://doi.org/10.1126/science.aeb1178

Hester WP (2019) *UVA Plans New School of Data Science; $120 Million Gift is Largest in University History.* https://archive.news.virginia.edu/content/uva-plans-new-school-data-science-120-million-gift-largest-university-history (11 August 2026, date last accessed).

ISCB-News-Mar2026 (International Society for Computational Biology). *March 2026 Newsletter*. https://www.iscb.org/about-iscb/society-communications/announcements/march-30-2026-iscbs-march-2026-newsletter (11 August 2026, date last accessed).

Johnson TR, Bourne PE. The Biological Data Sustainability Paradox. *arXiv [q-bio.OT].* 2023; http://dx.doi.org/10.48550/arXiv.2311.05668

Jumper J, Evans R, Pritzel A *et al*. Highly accurate protein structure prediction with AlphaFold. *Nature.* 2021; **596**(7873): 583–589. https://doi.org/10.1038/s41586-021-03819-2

Kittrie E, Atienza AA, Kiley R *et al*. Developing international open science collaborations: Funder reflections on the Open Science Prize. *PLOS Biol.* 2017; **15**(8): e2002617. https://doi.org/10.1371/journal.pbio.2002617

Masum H, Bourne PE. Ten simple rules for humane data science. *PLOS Comput. Biol.* 2023; **19**: e1011698. https://doi.org/10.1371/journal.pcbi.1011698

Mather M. (2026) *Board gives final OK for two large projects that will alter Grounds' landscape.* https://news.virginia.edu/content/board-gives-final-ok-2-large-projects-will-alter-grounds-landscape (11 August 2026, date last accessed).

McKiernan EC, Bourne PE, Brown CT *et al*. How open science helps researchers succeed. *eLife*. 2016; **5**: e16800. https://doi.org/10.7554/eLife.16800

Microsoft-News-2010. *Jim Gray eScience Award: 2010 recipient.* https://www.microsoft.com/en-us/research/wp-content/uploads/2016/09/jimgrayaward.pdf (13 August 2026, date last accessed).

Moody M (2019) *UVA Board Approves Establishment of School of Data Science.* https://archive.news.virginia.edu/content/uva-board-approves-establishment-school-data-science (11 August 2026, date last accessed).

Mura C, Candelier E, Xie L. A Tribute to Phil Bourne—Scientist and Human. *Biomolecules*. 2023; **13**(1): 181. https://doi.org/10.3390/biom13010181

Mura C, Veretnik S, Bourne PE. The Urfold: Structural similarity just above the superfold level? *Protein Sci*. 2019; **28**(12): 2119–2126. https://doi.org/10.1002/pro.3742

Ono K. (2026) *Comments on Phil Bourne's passing.* https://www.linkedin.com/posts/ken-ono-a972191a5_i-have-been-struggling-to-process-news-about-activity-7438243219408171008-h92k (11 August 2026, date last accessed).

Peters B, Sidney J, Bourne PE *et al*. The immune epitope database and analysis resource: from vision to blueprint. *PLOS Biol*. 2005; **3**(3): e91. https://doi.org/10.1371/journal.pbio.0030091

Ponomarenko J, Bui HH, Li W *et al*. ElliPro: a new structure-based tool for the prediction of antibody epitopes. *BMC Bioinformatics*. 2008; **9**(1): 514. https://doi.org/10.1186/1471-2105-9-514

Ponomarenko J, Papangelopoulos N, Zajonc DM, *et al*. IEDB-3D: Structural data within the immune epitope database. *Nucleic Acids Res*. 2011; **39**: D1164–70. https://doi.org/10.1093/nar/gkq888

Puebla I. (2026) *A Tribute to Phil Bourne – FORCE11.* https://force11.org/post/a-tribute-to-phil-bourne (13 August 2026, date last accessed).

Scheeff ED, Bourne PE. Structural evolution of the protein kinase-like superfamily. *PLoS Comput. Biol*. 2005; **1**: e49. https://doi.org/10.1371/journal.pcbi.0010049

SDSC-News-Apr2009 (San Diego Supercomputer Center). *Open Access Advocate Philip E. Bourne to Receive 2009 Benjamin Franklin Award.* https://www.sdsc.edu/news/2009/PR042809_franklinawa.html (11 August 2026, date last accessed).

Seeland U, Coluzzi F, Simmaco M *et al*. Evidence for treatment with estradiol for women with SARS-CoV-2 infection. *BMC Med*. 2020; **18**(1): 369. https://doi.org/10.1186/s12916-020-01851-z

Shindyalov IN, Bourne PE. An alternative view of protein fold space. *Proteins.* 2000; **38**(3), 247–260. https://doi.org/10.1002/(SICI)1097-0134(20000215)38:3%3C247::AID-PROT2%3E3.0.CO;2-T

Shindyalov IN, Bourne PE. Protein structure alignment by incremental combinatorial extension (CE) of the optimal path. *Protein Eng*. 1998; **11**(9): 739–747. https://doi.org/10.1093/protein/11.9.739

UVA-DataSci-Memorial. (2026) *Founding Dean Bourne Memorial.* https://youtu.be/-m2q1q5oucs?si=a1AKPMM36lpsZlpu (12 August 2026, date last accessed).

UVA-DataSci-Story. (2026) *The Story of Us*. https://story.datascience.virginia.edu/chapters/a-growing-school (12 August 2026, date last accessed).

Veretnik S, Wills C, Youkharibache P *et al*. Sm/Lsm genes provide a glimpse into the early evolution of the spliceosome. *PLoS Comput. Biol.* 2009; **5**(3): e1000315. https://doi.org/10.1371/journal.pcbi.1000315

Westbrook JD, Bourne PE. STAR/mmCIF: An ontology for macromolecular structure. *Bioinformatics*. 2000; **16**(2): 159–168. https://doi.org/10.1093/bioinformatics/16.2.159

Wilkinson MD, Dumontier M, Aalbersberg IJ *et al*. The FAIR Guiding Principles for scientific data management and stewardship. *Sci. Data*. 2016; **3**(1): 160018. https://doi.org/10.1038/sdata.2016.18

Wiper ML. The 2026 ISCB Outstanding Service Award—Dr Philip E. Bourne. *Bioinformatics*. 2026; **42**, btag280. https://doi.org/10.1093/bioinformatics/btag280

Xie L, Bourne PE. Detecting evolutionary relationships across existing fold space, using sequence order-independent profile-profile alignments. *Proc. Natl. Acad. Sci. USA*. 2008; **105**(14), 5441–5446. https://doi.org/10.1073/pnas.0704422105

Zhao Z, Bourne PE. How ligands interact with the kinase hinge. *ACS Med. Chem. Lett*. 2023; **14**: 1503–1508. https://doi.org/10.1021/acsmedchemlett.3c00212

Zhao Z, Bourne PE. Deciphering covalent kinase inhibitor binding landscape through structural kinome profiling. *Eur. J. Med. Chem*. 2026; **312**: 118872. https://doi.org/10.1016/j.ejmech.2026.118872